\documentclass[12pt, draftclsnofoot, onecolumn]{IEEEtran}

\normalsize

\usepackage{amsmath, amsfonts, amssymb, amsthm}
\usepackage{bbm}
\usepackage{graphicx}
\usepackage{epstopdf}
\usepackage[noabbrev]{cleveref}
\usepackage{scalefnt}
\usepackage{indentfirst}
\usepackage{cite}
\usepackage{makecell}
\usepackage[caption=false,subrefformat=parens,labelformat=parens]{subfig}

\usepackage{algorithm}
\usepackage{algcompatible}
\usepackage{multirow}

\begin{document}

{\Large \textbf{Notice:} This work has been submitted to the IEEE for possible publication. Copyright may be transferred without notice, after which this version may no longer be accessible.}
\clearpage

\title{Sparse Vector Transmission: An Idea Whose Time Has Come}

\author{
Wonjun Kim$^{\dagger}$, Hyoungju Ji$^{\ast}$, Hyojin Lee$^{\ast}$, Younsun Kim$^{\ast}$, Juho Lee$^{\ast}$, and Byonghyo Shim$^{\dagger}$ \\
$^{\dagger}$Institute of New Media and Communications and Department of Electrical and Computer Engineering, Seoul National University, Seoul, Korea \\
$^{\ast}$Samsung Research Seoul R\&D Center, Seoul, Korea \\
\thanks{This work was supported by 'The Cross-Ministry Giga KOREA Project' grant funded by the Korea government(MSIT) (No. GK18P0500, Development of Ultra Low-Latency Radio Access Technologies for 5G URLLC Service).}
}

\maketitle

\begin{abstract}
In recent years, we are witnessing bewildering variety of automated services and applications of vehicles, robots, sensors, and machines powered by the artificial intelligence technologies. Communication mechanism associated with these services is dearly distinct from human-centric communications. One important feature for the machine-centric communications is that the amount of information to be transmitted is tiny. In view of the short packet transmission, relying on today's transmission mechanism would not be efficient due to the waste of resources, large decoding latency, and expensive operational cost. In this article, we present an overview of the sparse vector transmission (SVT), a scheme to transmit a short-sized information after the sparse transformation. We discuss basics of SVT, two distinct SVT strategies, viz., frequency-domain sparse transmission and sparse vector coding with detailed operations, and also demonstrate the effectiveness in realistic wireless environments.
\end{abstract}
 
\newpage

\section{Introduction}
These days, automated things such as vehicles, drones, sensors, machines, and robots, combined with artificial intelligence (AI) technologies, have found their way into almost every industry.
Remarkable growth of business models using autonomous machines is accelerating the need for communication between machines as well as machine to human communications~\cite{Taleb_Machine}.
One important feature of machine-centric communications over the long-standing human-oriented communications is that the amount of information to be transmitted is tiny.
For example, information to be exchanged in the autonomous driving, robot, smart factory, and home appliance is in a form of control and command-type information such as start/stop, turn on/off, move left/right, speed up/slow down, shift, and rotate.
Typically, required information bit in these applications is in the range of 10 $\sim$ 100 bits.
Information acquired from the sensors (e.g., temperature, pressure, speed, gas density) is in the order of 10 bits.
Also, similar sized packets are used in many feedback or control channels (e.g., ACK/NACK feedback in 4G LTE/5G NR PUCCH~\cite{Sesia_LTE, 3GPP_Technical}).

Crucial observation in these applications is that conventional transmission mechanism is unduly complicated and inefficient, resulting in a waste of resources, transmit power, and processing latency.
Shannon's channel coding theorem, governing principle of today’s packet transmission, is based on the law of large numbers so that it works properly only when the packet size is sufficiently large.
In fact, when the packet length is short, noise introduced by the channel cannot be averaged out properly, degrading the packet reception quality substantially (see, e.g., information theoretic analysis in~\cite{Polyanskiy_Channel}).
Further, in the ultra short-packet transmission regime, size of the non-payload (pilot signals and control data) easily exceeds the payload size so that the cost caused by the non-payload outweighs the cost of payload.
In particular, in some applications requiring high reliability (e.g., ultra-reliable and low latency communications (URLLC) in 5G~\cite{Ji_Ultra}), cost caused by the pilot signaling increases sharply, further degrading the resource utilization efficiency.
Without doubt, relying on today's transmission mechanism would not be efficient due to the waste of resources, large decoding latency, and also expensive operational cost.

Our intent in this article is to introduce new type of short packet transmission scheme referred to as sparse vector transmission (SVT).
Key idea of SVT is to transmit the short-sized information after the sparse vector transformation.
Using the principle of compressed sensing (CS), we decode the packet using a small number of resources.
SVT has a number of advantages over the conventional transmission strategies; it is simple to implement, reduces the transmission latency as well as the encoding/decoding complexity.
When the position of a sparse vector is used to encode the information exclusively, decoding can be done without the channel knowledge, saving the pilot transmission overhead and the channel estimation effort.
Further, SVT can inherently improve the user identification quality and security.
In a nutshell, SVT is a viable solution for massive machine-type communication (mMTC) and URLLC scenarios having many advantages over the conventional packet transmission mechanism.

The rest of this paper is organized as follows.
In Section II, we explain the principle of sparse vector transmission.
In Section III and IV, we discuss two distinct SVT strategies: the frequency domain sparse transmission and sparse vector coding.
We present the future research direction and conclusion in Section V.

\section{Principle of Sparse Vector Transmission}
In this section, we discuss the basic principle of SVT.
After the brief review of the CS principle, we explain two types of SVT schemes: \textit{frequency-domain sparse transmission} and \textit{sparse vector coding}.

\subsection{Basics of Compressed Sensing}
A system with $m$-dimensional measurement vector $\mathbf{y}$ and $n$-dimensional input vector $\mathbf{s}$ given by
$$\mathbf{y}=\mathbf{A}\mathbf{s}$$
where $\mathbf{A}$ is the system (sensing) matrix relating the input vector $\mathbf{s}$ and the measurement vector $\mathbf{y}$.
If $\mathbf{A}$ is a tall or square matrix, meaning that the dimension of $\mathbf{y}$ is larger than or equal to the dimension of $\mathbf{s}$ $(m \geq n)$, one can recover $\mathbf{s}$ using the conventional techniques (e.g., Gaussian elimination) as long as the sensing matrix is a full rank.
However, if the matrix $\mathbf{A}$ is fat $(m < n)$, meaning that the number of unknowns is larger than the dimension of observation vector, it is in general not possible to find out the unique solution since there exist infinitely many possible solutions.
In this pathological scenario where the inverse problem is ill-posed, sparsity of an input vector comes to the rescue.
A vector $\mathbf{s}$ is called sparse if the number of nonzero entries is sufficiently smaller than the dimension of the vector.
If a vector $\mathbf{s}$ is $k$-sparse, meaning that there are $k$ nonzero elements in $\mathbf{s}$, the measurement vector $\mathbf{y}$ is expressed as a linear combination of $k$ columns of $\mathbf{A}$ associated with the nonzero entries of $\mathbf{s}$.
If the support $\Omega_{\mathbf{s}}$ (set of nonzero indices in $\mathbf{s}$) is known a priori by any chance, then by removing columns corresponding to the zero entries in $\mathbf{s}$, we can convert the underdetermined system into over-determined one and thus can find out the solution using the standard technique.
CS theory asserts that as long as the sensing matrix is generated at random, $k$-sparse vector can be recovered with $m \approx ck\log(n/k)$ measurements ($c$ is a constant).
In performing the recovery task, $\ell_{1}$-norm minimization technique and greedy sparse recovery algorithm (e.g., orthogonal matching pursuit (OMP)) have been popularly used (see, e.g.,~\cite{CS_Trick}).

It is worth mentioning that underlying assumption in many CS-based studies is that the signal is sparse in nature or can be sparsely represented in a properly chosen basis.
Indeed, applications of CS in wireless communications have been mainly on the recovery of naturally sparse signals such as sparse millimetre wave channel estimation in angle and delay domains, angle (DoA/AoD) estimation, and spectrum sensing~\cite{CS_Trick, Gao_Compressive, Liao_Closed}.
Intriguing feature of SVT over these is to \textit{purposely} transmit the sparse vector to achieve the gain in performance, latency, and energy efficiency.

\begin{figure}
	\centering
	\includegraphics[width=.65\columnwidth]{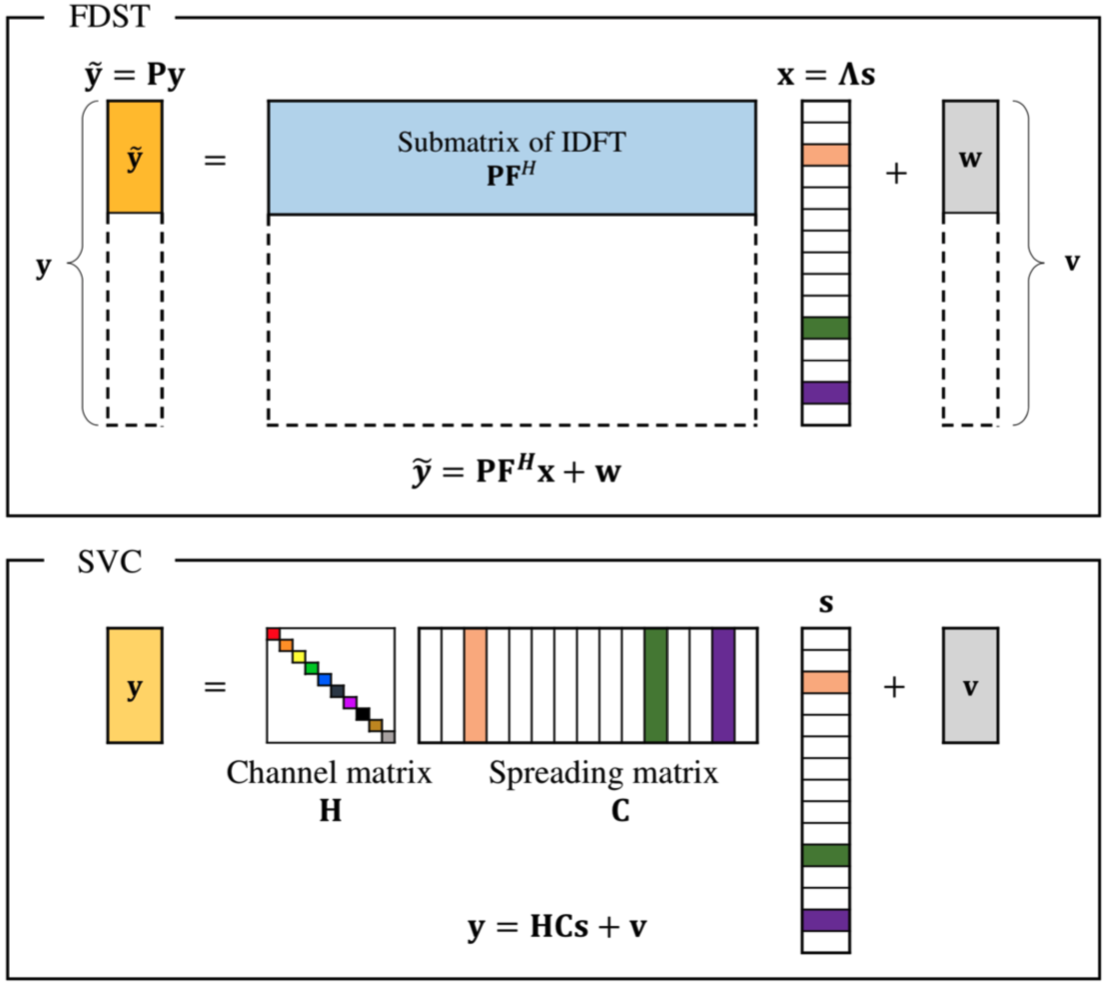}
	\caption{System models for SVT schemes (FDST and SVC).}
	\label{fig_1}
\end{figure}

\subsection{Two Types of SVT}
Basically, there are two options in SVT.
To ease our exposition, we discuss the orthogonal frequency division multiplexing (OFDM) system, standard systems for 4G, 5G cellular and WiFi systems, as a baseline.
Nevertheless, main principle can be readily extended to different transmission schemes.
In the first option, referred to as the frequency-domain sparse transmission (FDST), information is embedded into a small number of subcarriers and then transmitted (see Figure 1).
In this case, composite of the channel matrix $\mathbf{H}$ and the IDFT matrix $\mathbf{F}^{H}$ becomes the sensing matrix so that the time-domain sample vector $\mathbf{y}=\mathbf{A}\mathbf{s}+\mathbf{v}=\mathbf{H}\mathbf{F}^{H}\mathbf{s}+\mathbf{v}$ becomes the measurement vector.
While the symbol decoding in the conventional OFDM systems is initiated after receiving all time-domain samples, sparse vector $\mathbf{s}$ in FDST can be recovered with a small number of time-domain samples using the CS technique.
Let $\mathbf{P}$ be the matrix taking early $m$ samples of $\mathbf{y}$, then the vector of the first $m$ measurements is expressed as $\tilde{\mathbf{y}}=\mathbf{P}\mathbf{y}$ (see Figure 1).
For instance, if $k=16$, $n=1024$, and $c=4$, then only 11\% of samples $(m \approx 115)$ is needed to decode $\mathbf{s}$.
In the second option, called the sparse vector coding (SVC), spreading matrix $\mathbf{C}=[\mathbf{c}_{1} \ \cdots \ \mathbf{c}_{n}]$ is applied to the sparse vector $\mathbf{s}$ before the transmission (see Figure 1).
Each position in a vector $\mathbf{s}$ has its own spreading sequence $\mathbf{c}_{i}$ so that the transmit vector can be expressed as a linear combination of spreading sequences corresponding to nonzero symbols.

Distinctive feature of SVT over the conventional transmission scheme is that positions as well as symbols can be employed to convey the information.
By way of analogy, one can imagine a process to generate the sparse vector as placing a few balls into the empty boxes.
When we try to put $k$ balls in $n$ boxes $(k \leq n)$, we have ${n \choose k}$ choices, so that we can encode $\lfloor \log_{2} {n \choose k} \rfloor$ bits of information into the position of the sparse vector $\mathbf{s}$.
Suppose the modulation order is the same for all nonzero positions (say $b_{s}$ bit per symbol), then $kb_s$ bits can be encoded to the active symbols (symbols in the nonzero positions) so that one SVT block conveys $kb_s + \lfloor\log_{2} {n \choose k}\rfloor$ bits in total.

There are various options to encode the information in SVT.
One simple option is to use both positions and active symbols in the information transmission.
Alternatively, one can map the user ID (UID) to the positions and the rest information to the active symbols to elegantly divide the user identification process and information decoding.
Yet another option is to embed the message to the positions and the parity bits for the error detection and correction to the active symbols.

\section{Frequency-domain Sparse Transmission}
In this section, we discuss the FDST scheme in detail.
In contrast to the conventional OFDM systems, FDST transmits the information in a form of a sparse vector and then uses the CS technique to decode the input sparse vector.
We first discuss the system model and then explain the FDST decoding and environment-aware user identification, an approach to simplify the user identification process using environmental information.

\subsection{System Model}
As discussed, the system model for FDST is $\mathbf{y}=\mathbf{H}\mathbf{F}^{H}\mathbf{s}+\mathbf{v}$.
Due to the addition of the cyclic prefix, $\mathbf{H}$ is a circulant matrix and thus can be eigen-decomposed by DFT basis.
That is, $\mathbf{H}=\mathbf{F}^{H}\boldsymbol{\Lambda}\mathbf{F}$ where $\mathbf{F}$ is the DFT matrix and $\boldsymbol{\Lambda}$ is the diagonal matrix ($\lambda_{ii}$ represents the channel of $i$-th subcarrier).
The corresponding system model is expressed as $\mathbf{y}=\mathbf{F}^{H}\boldsymbol{\Lambda}\mathbf{s}+\mathbf{v}$.
Since the supports of $\mathbf{s}$ and $\boldsymbol{\Lambda}\mathbf{s}$ are the same, by letting $\mathbf{x}=\boldsymbol{\Lambda}\mathbf{s}$, the system model is converted to $\mathbf{y}=\mathbf{F}^{H}\mathbf{x}+\mathbf{v}$.
Recalling that CS operates with far fewer measurements than the conventional techniques require, a small part of $\mathbf{y}$ is enough to recover $\mathbf{x}$.
From the principle of CS, the recovery performance depends heavily on the quality of sensing matrix.
As a metric to evaluate the sensing matrix, mutual coherence defined as the largest magnitude of normalized inner product between two distinct columns of sensing matrix is widely used~\cite{CS_Trick}.
As long as we choose consecutive samples, the mutual coherence of IDFT submatrix $(\mathbf{P}\mathbf{F}^{H})$ remains constant so that the recovery performance would not be affected by the choice of samples in $\mathbf{y}$.
Hence, one can use early arrived samples to achieve a reduction in transmission and decoding latencies (see Figure 2).
Since this is a standard setting for CS, any sparse recovery algorithm can be employed to decode $\mathbf{x}$ from $\tilde{\mathbf{y}}$.

\subsection{FDST Decoding}
Basically, decoding of FDST consists of two steps; in the first step, support of $\mathbf{x}$ is identified by the sparse recovery algorithm.
For example, greedy sparse recovery algorithm identifies one column (position of a vector) of the sensing matrix in each iteration.
In our case, a column of $\mathbf{P}\mathbf{F}^{H}$ that is maximally correlated with the measurement vector is chosen.
Once the support $\Omega_{\mathbf{x}}$ of $\mathbf{x}$ (equivalently the support $\Omega_{\mathbf{s}}$ of $\mathbf{s}$) is identified, by removing components associated with the zero entries in $\mathbf{s}$, an over-determined system model to decode the symbol $\mathbf{s}$ can be obtained.
In the decoding of $\mathbf{s}$, conventional technique such as the linear minimum mean square error (LMMSE) estimator followed by the symbol slicer can be used for the symbol detection.

\begin{figure}
	\centering
	\includegraphics[width=.9\columnwidth]{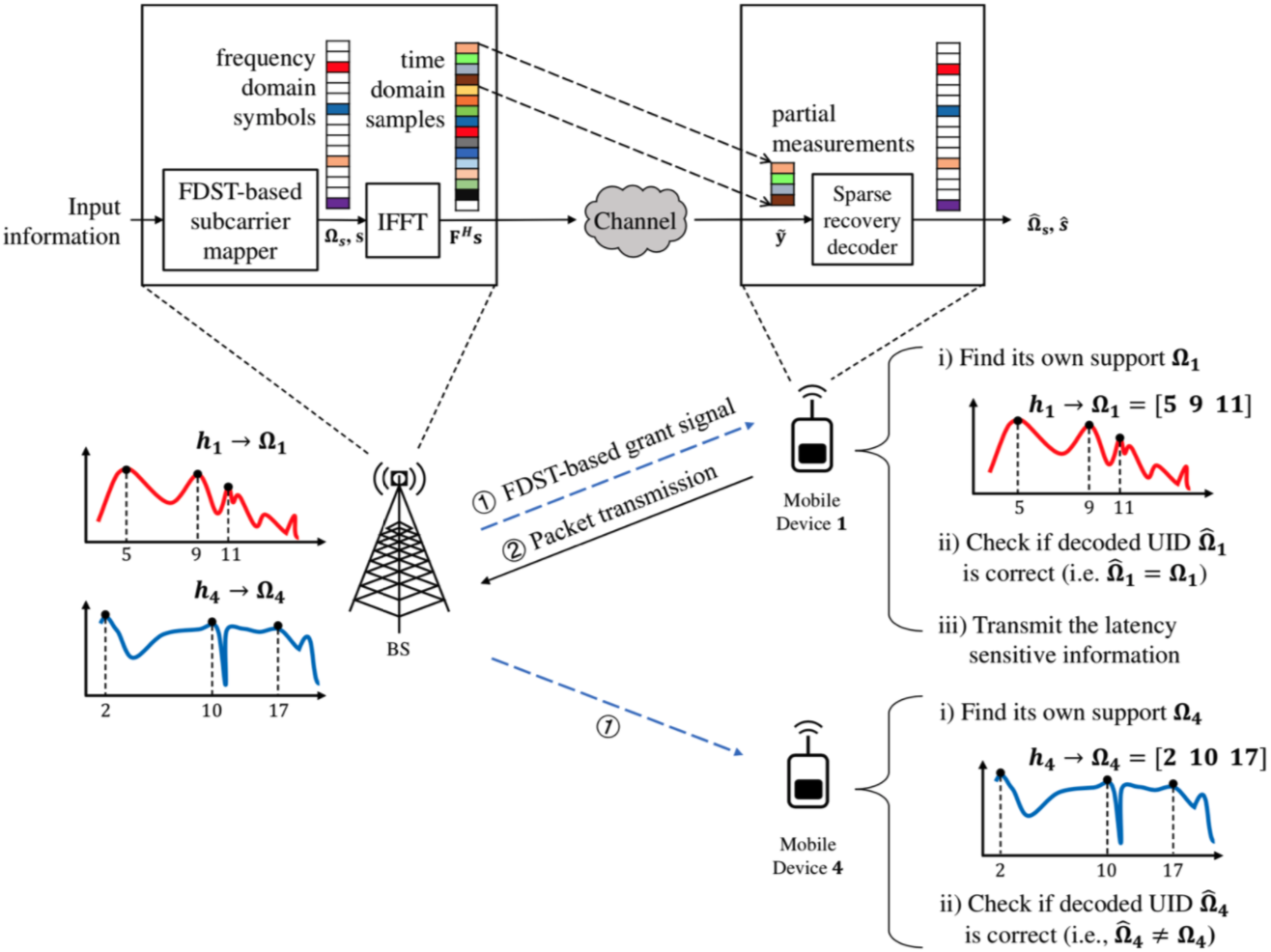}
	\caption{Illustration of the FDST-based short packet transmission in the TDD systems $(k=3)$.}
	\label{fig_2}
\end{figure}

In the mapping of the UID to the support, one can consider the environment-aware user identification (EA-UI).
Key idea of EA-UI is to use a support $\Omega_{\mathbf{s}}$ derived from the environmental information as a UID.
By the environmental information, we mean the information obtained from wireless environments such as channel impulse response (CIR), angle (AoD, DoA), location, delay spread, to name just a few.
EA-UI is conceptually similar to the biometric user identification.
Biometric identifier, such as iris or fingerprint, intrinsically representing the unique identity of individual’s body, can greatly simplify the user identification process.
Principle of EA-UI is essentially the same since the environmental information is reflected in the support of the transmit vector.
One simple example, illustrated in Figure 2, is to choose positions of $k$ subcarriers having the largest channel gain as a support and use this as a UID.
For example, in TDD systems, base station (BS) can acquire the channel information (and thus UID) of all mobile devices from the uplink pilot signals due to the channel reciprocity.
In the uplink scenario, therefore, BS can identify which mobile device has sent the packet by checking the support (UID) of the received packet.
Similarly, in the downlink scenario, mobile device can easily check whether the packet is for itself by comparing the decoded support $\hat{\Omega}_{\mathbf{x}}$ and its own support $\Omega_{\mathbf{x}}$.

EA-UI has several advantages; first, it improves the security since the UID is derived from naturally acquired environmental (channel) information.
Second, in most physical channels (e.g., UE-specific channels like PDCCH, PDSCH in 4G LTE/5G NR~\cite{3GPP_Technical}), BS sends the data together with UID to notify which device the packet is delivered to.
Since the EA-UI mechanism separates the user identification and data decoding elegantly, time and effort to decode whole packet just for the user identification purpose can be saved.
Indeed, since EA-UI is done by the identification of the support in $\mathbf{x}$, not by the accurate recovery of sparse vector $\mathbf{s}$, support of $\mathbf{x}$ can be recovered using $\tilde{\mathbf{y}}$ and $\mathbf{D}=\mathbf{P}\mathbf{F}^{H}$ (the submatrix of IDFT).
Recalling that $\mathbf{D}$ is independent of the channel and $\mathbf{x}=\boldsymbol{\Lambda}\mathbf{s}$, the channel estimation is unnecessary in the support detection.
Third, since $k$ is in general very small, sparse recovery algorithm can quickly identify the support.
Further, transmission and decoding latency can be greatly reduced since only a small fraction of early arrived (time-domain) samples is used for the packet decoding (see Figure 2).

As a final note, one can easily add the error correction capability to EA-UI since the sparsity of subcarriers lends itself to the addition of error correction mechanism.
For example, when the correlation between adjacent columns in $\mathbf{D}$ is large, which is true for the submatrix of IDFT, an index of a column adjacent to the correct one can be chosen as a support element incorrectly.
Incorrect support element can also be chosen when the supports of BS and UE are slightly different due to the channel estimation error or imperfect channel reciprocity.
Since
$k$ is small $(k \ll n)$ in $\mathbf{s}$, by relaxing the success condition in the support identification, an error can be corrected.
Basic idea of this strategy is to replace the selected index $\hat{\omega}$ with the nearest support element $\omega \in \Omega_{\mathbf{x}}$.
In other words, as long as the mismatch level $\hat{\omega}-\omega$ is smaller than the properly designed threshold, the error caused by the different supports can be corrected~\cite{Kim_Channel}.

\begin{table}[t]
	\centering
	\caption{System Setup for FDST Simulations}
	\begin{tabular}{|l||c||c|l|}
		\hline
		& FDST  & \multicolumn{2}{c|}{PDCCH with convolution and turbo code (1/3 rate)} \\ \hline \hline
		System model & \multicolumn{3}{c|}{5 MHz bandwidth, 15 kHz spacing, and 1 subframe = 1 ms}                \\ \hline
		FFT size & \multicolumn{3}{c|}{512 ($k=36$ in FDST)}                                                 \\ \hline
		Channel model & \multicolumn{3}{c|}{i.i.d Rayleigh fading channel}                                         \\ \hline
		Number of bit & 144 & \multicolumn{2}{c|}{160 (144 for control information and 16 for UID)}    \\ \hline
		Modulation scheme & 16-QAM & \multicolumn{2}{c|}{QPSK}                                                   \\ \hline
	\end{tabular}
\end{table}

\begin{figure}
	\centering
	\includegraphics[width=.6\columnwidth]{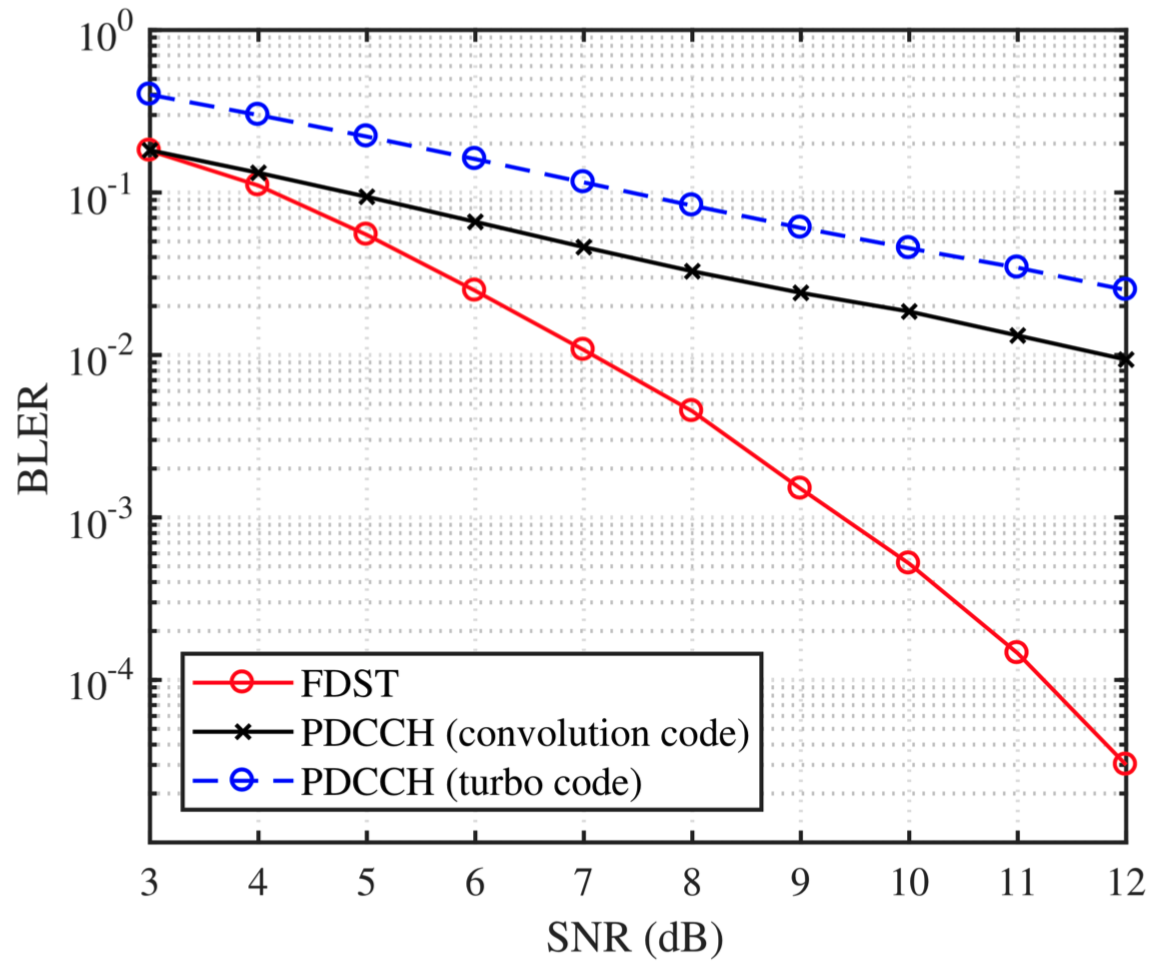}
	\caption{BLER of URLLC packet transmission $(m=256)$.}
	\label{fig_3}
\end{figure}

\subsection{Numerical Performance Evaluation}
In this subsection, we present the numerical results to evaluate the performance of FDST.
In our simulations, the OFDM systems (with 512 subcarriers) under the i.i.d. Rayleigh fading channels are used.
As performance measures, the block error rate (BLER) and the processing latency (see Table I for detailed setup) are considered.
In Figure 3, the BLER of FDST and PDCCH in 4G LTE are compared.
In our simulations, sizes of payload and non-payload are set to 144 and 16 bits, respectively.
Due to the selective use of good subchannels and also properly designed error correction mechanism, FDST outperforms PDCCH by a large margin, achieving more than 5 dB gain when BLER is $10^{-2}$.
We next evaluate the average processing latency defined as the sum of the buffering latency and decoding latency for one OFDM symbol.
The processing latencies of FDST for $m=256$ $(73.4 \mu s)$ and $m=128$ $(36.7 \mu s)$ are reduced by the factor of 56\% and 78\% over the LTE PDCCH $(166.8 \mu s)$, respectively.
In practice, to decode a packet in 4G LTE/5G NR systems, we need to receive 7 (4G LTE) or 2 (5G NR) OFDM symbols while only one symbol is enough for the proposed FDST so that the latency gain of FDST is pronounced.
We note that, since the required number of samples in the receiver is small, the base station does not need to transmit whole samples, resulting in the saving of the transmit power.

\section{Sparse Vector Coding}
In this section, we discuss the SVC transmission scheme.
Distinctive feature of SVC over FDST is to transmit the sparse vector after the random spreading.
We first discuss the basic SVC operation and then explain the deep neural network based SVC decoding.

\begin{figure}
	\centering
	\includegraphics[width=.6\columnwidth]{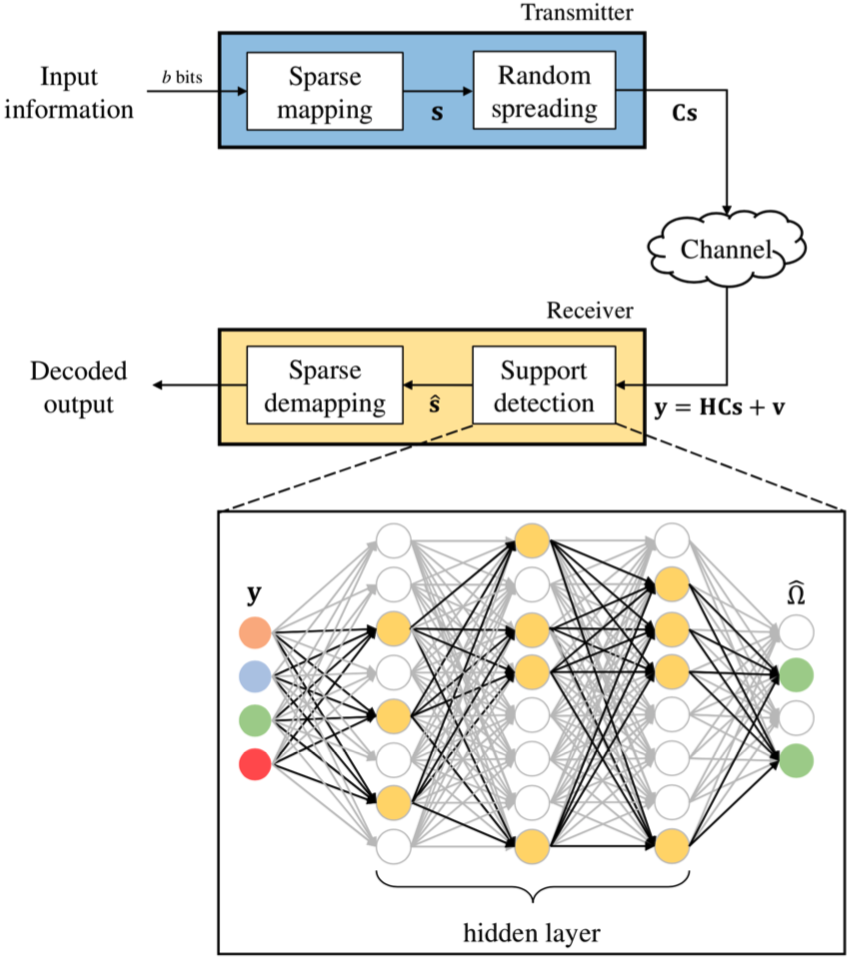}
	\caption{The block diagram for the SVC transceiver using the deep neural network.}
	\label{fig_4}
\end{figure}

\subsection{Basics of SVC}
Key idea behind SVC is to transmit the short-sized packet after the sparse vector transformation and compression.
To be specific, an information vector is mapped to the sparse vector and then transmitted after the pseudo random spreading (see Figure 4).
In view of this, one can consider SVC as a reminiscent of CDMA.
However, major difference is that the length of the spreading sequence $\mathbf{c}_{i}$ is far shorter than the size of an input vector $\mathbf{s}$ and only a few entries in $\mathbf{s}$ are nonzero.
Note, in contrast to FDST using the sub-matrix of IDFT as a sensing matrix, the sensing matrix $\mathbf{C}=[\mathbf{c}_{1} \ \mathbf{c}_{2} \ \cdots \ \mathbf{c}_{n}]$ of SVC is generated at random so that the vector $\mathbf{C}$$\mathbf{s}$ after the spreading contains enough information to recover the sparse vector $\mathbf{s}$.
The input-output relationship of SVC is $\mathbf{y}=\mathbf{H}\mathbf{C}\mathbf{s}+\mathbf{v}$ where $\mathbf{H}$ is the diagonal matrix where $h_{ii}$ is the channel for $i$-th subcarrier (e.g., resource element in LTE) and $\mathbf{v}$ is the additive Gaussian noise.

Major benefit of SVC is that the decoding process can be greatly simplified when the channel is approximately constant, which is true for most short packet transmission scenarios.
Note that when the channel is approximately constant (i.e., $\mathbf{H} \approx h\mathbf{I}$), the system model can be simplified to $\mathbf{y} \approx h\mathbf{C}\mathbf{s}+\mathbf{v}=\mathbf{C}\mathbf{x}+\mathbf{v}$ where $\mathbf{x}=h\mathbf{s}$.

Validity of this system model can be fortified when the packet is transmitted in a narrowband channel or approximately static environments.
In fact, when the packet transmission time $T_{p}$ is smaller than the channel coherence time $T_{c}$ $(T_c \gg T_p)$, channel can be readily assumed to be a constant.
For example, when the carrier frequency $f_c=1.8$ GHz and the mobile speed is $\nu=10$ km/h, then $T_c \approx 11$ ms is much larger than the duration of one LTE OFDM symbol $(\approx 0.07 \text{ ms})$~\cite{Tse_Fundamentals}, which justifies the validity of this approximation.

Suppose an information is embedded only in the positions of $\mathbf{s}$.
Then the decoding task is to find out the support $\Omega_{\mathbf{s}}$ of $\mathbf{s}$ (equivalently $\Omega_{\mathbf{x}}$).
Fortunately, since the system matrix $\mathbf{C}$ does not contain the channel components, the support identification can be done without the channel knowledge, resulting in savings of the resources and power for the pilot transmission, not to mention the saving the trouble for the channel estimation.
Also, SVC is also good fit for the small cell deployed scenarios or cell-edge environments since the randomly spread sparse vector is robust to the co-channel interference.

\subsection{SVC Decoding with Deep Neural Network}
In the SVC packet decoding, we can basically use any sparse signal recovery algorithm.
In most sparse recovery algorithms, such as OMP, an index of a column in $\mathbf{C}$ that is maximally correlated to the measurement $\mathbf{y}$ is chosen as an estimate of the support element~\cite{Pati_Orthogonal, Wang_Generalized}.
Thus, if two columns of $\mathbf{C}$ are strongly correlated and only one of these is associated with the nonzero values in $\mathbf{s}$, then it might not be easy to distinguish the right column from wrong one.
Clearly, support identification performance depends heavily on the correlation between columns of $\mathbf{C}$.
In fact, when a packet is transmitted using a small amount of resources, underdetermined ratio $m/n$ of the system will increase sharply, causing a severe degradation in the decoding performance.

To address this problem, we employ a deep neural network (DNN), an outstanding tool to approximate the complicated and nonlinear function.
Particularly, in this work, we use a convolutional neural network (CNN) due to its computation efficiency~\cite{Krizhevsky_Imagenet}.
In the training phase, CNN learns the nonlinear mapping function $g$ between the received signal $\mathbf{y}$ and support $\Omega$ (see Figure 4).
The support identification problem can be expressed as $\hat{\Omega}=g(\mathbf{y};\boldsymbol{\Theta})$ where $\boldsymbol{\Theta}$ is the set of weights and biases.
In order to find out the network parameters $\boldsymbol{\Theta}$ that approximate the mapping function correctly, a received signal $\mathbf{y}$ passes through multiple hidden layers.
In each hidden layer, the relationship between input $\mathbf{z}^{i}$ and output $\mathbf{z}^{o}$ can be expressed as $\mathbf{z}^{o}=f(\mathbf{W} \ast \mathbf{z}^{i} + \mathbf{b})$ where $\mathbf{W}$ and $\mathbf{b}$ are the weight and bias, respectively, $f$ is the nonlinear activation function (e.g., ReLu function), and $\ast$ is the convolution operator.
In order to achieve a reduction in the computational complexity, the max pooling operation taking the maximal value in the filtered window is employed.
It is now well-known from the universal approximation theorem that DNN processed by the deeply stacked hidden layers well approximates the desired function.
In our context, this means that the trained neural network with multiple hidden layers can handle the whole SVC decoding process, resulting in an accurate support identification.

When one tries to use the deep learning techniques to the wireless communication systems (e.g., packet decoding and channel estimation), there are two major problems in the training phase.
First, it is very difficult to handle the randomness of the received vector $\mathbf{y}$.
Indeed, since the channel depends heavily on the environmental factors such as frequency band and geometric objects, DNN needs to learn a huge amount of channel training data.
Fortunately, the SVC decoding is essentially the same as the support identification and all channel components are contained in an input sparse vector $\mathbf{x}=h\mathbf{s}$.
Thus, the DNN-based SVC decoding only needs to learn the codebook matrix $\mathbf{C}$ (known in advance), not the channel statistics, which greatly simplifies the learning process and also improves accuracy.

Next, in the training phase, abundant amount of dataset is required to train the DNN.
Using the real received signals as training data would be a natural option but it requires huge training overhead.
For example, when gathering ten million received signals in LTE systems, it will take more than 4 hours (1 ms subframe consisting of 14 symbols).
Luckily, since the DNN-based SVC decoding is not so sensitive to the channel variations, the training vectors can be generated synthetically and then used in the offline training phase.
In doing so, time and effort to collect huge training data can be prevented.

\begin{figure}
	\centering
	\includegraphics[width=.6\columnwidth]{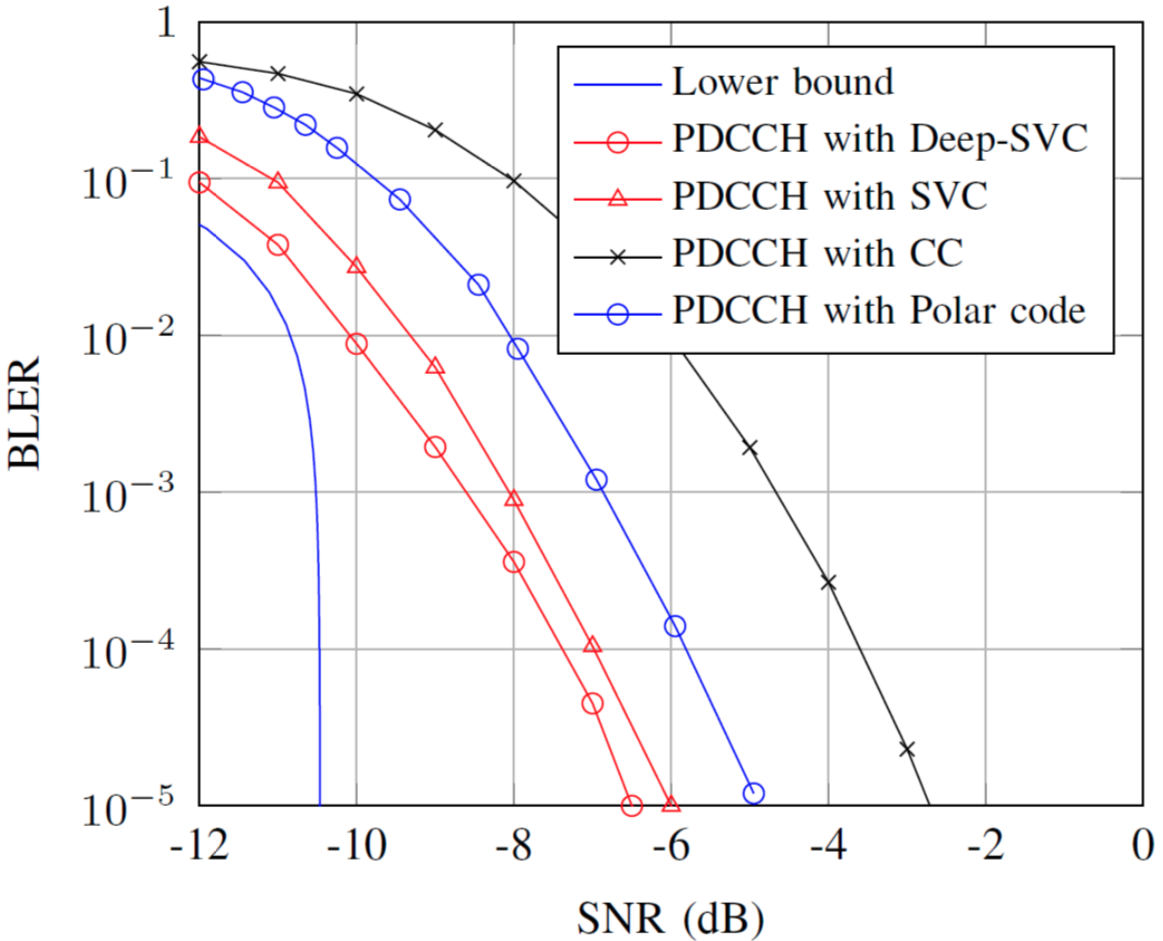}
	\caption{The block diagram for the SVC transceiver using the deep neural network.}
	\label{fig_5}
\end{figure}

\begin{table}[]
	\centering
	\caption{Computational complexity analysis for deep-SVC}
	\begin{tabular}{|c|l|c|c|}
		\hline
		\multirow{2}{*}{} & \multicolumn{1}{c|}{\multirow{2}{*}{The number of flops}}                                                                                                                   & \multicolumn{2}{c|}{Complexity for various input size} \\ \cline{3-4} 
		& \multicolumn{1}{c|}{}                                                                                                                                                       & $m=42$                     & $m=54$                    \\ \hline
		Deep-SVC          & $2mT\left(2q+\frac{p-1}{p}-1 + \frac{(4q+1)L}{p} - \frac{T-p}{T^{2}}\right) + (4m+k+3)n - \frac{k(k+1)}{2} -1$ & $1.36 \times 10^{5}$       & $1.75 \times 10^{5}$      \\ \hline
		OMP               & $2mnk + k(k+1)(4m^{2}+3m-1)/2 + km^{3}-2k$                                                                                                                                  & $1.86 \times 10^{5}$       & $3.71 \times 10^{5}$      \\ \hline
	\end{tabular}
\end{table}

\subsection{Numerical Performance Evaluation}
In this subsection, we examine the BLER performance of SVC-encoded short packet in the downlink of OFDM systems.
For comparison, we also test the PDCCH of 4G LTE systems and 5G NR system (using polar code) under AWGN channel condition.
In the 4G PDCCH, the convolution code with rate 1/3 with the 16-bit CRC is used.
In the SVC scheme, the binary codebook generated from the Bernoulli distribution is employed.
Also, we set $m=42$, $n=96$, and $k=2$ for 12 bits transmission.
In the decoding of the SVC-encoded packet, one-dimensional CNN consisting of 6 hidden layers whose size is 84 (twice as $m$) and convolution filter whose size is 3 are employed.
In Figure 5, we observe that the DNN-based SVC outperforms the conventional schemes, achieving 0.5 dB gain over the conventional SVC scheme, 3.3 dB gain over the PDCCH with convolutional code, and 1.2 dB gain over the PDCCH with polar code at BLER=$10^{-5}$.

In Table 2, we compare the computational complexities of the proposed Deep-SVC and conventional OMP algorithm under the simulation setup.
Here, the number of filters $T$ and pooling size $p$ are set to 32 and 2, respectively.
In order to examine the overall behaviour, we compute the required floating point operations (flops) for various input sizes $(m=42,54)$.
We observe that the complexity of Deep-SVC is much smaller than that of OMP.
For example, when $m=54$, the complexity of the proposed scheme is 53\% lower than that of the OMP algorithm.
It is worth mentioning that the complexity of the proposed scheme depends heavily on the DNN network parameter (e.g., $L$ and $T$), not the system parameter ($m$ and $k$).
For instance, when m increases from 42 to 54, the computational complexity of the Deep-SVC changes slightly but that of OMP algorithm increases significantly.

\section{Conclusion and Future Direction}
In this article, we presented an overview of the sparse vector transmission suitable for the short packet transmission in machine-centric communication scenarios (mMTC and URLLC).
We discussed basics of SVT, two distinct SVT approaches (FDST and SVC) with detailed operations, and also demonstrated the effectiveness of SVT in realistic wireless environments.
We observed that SVT is an effective means to transmit the short packet having many advantages over the conventional transmission scheme yet much work remains to be done.
For example, we did not elaborate the coding schemes in this work.
Perhaps simplest option is to combine the channel coding scheme and SVT mechanically.
Better option would be to consider the correlation of the sensing matrix and the quality of channel in the sparse vector generation and decoding.
In designing the coding scheme, we can recycle the wasted information (i.e., bit loss $\log_{2}{n \choose k} - \lfloor \log_{2}{n \choose k} \rfloor$).
Note that $b$ bits are mapped into $k$ positions of $n$-dimensional vector so that ${n \choose k}-2^{b}$ choices are wasted.
By the deliberate mapping of these choices to the information bits, decoding error probability can be reduced.
Also, SVT can simplify complicated transmission procedure.
For instance, SVT can be used as a grant signal in the user scheduling process.
It can also be used as a grant-free uplink transmission of the short packet.

As communication between machines proliferates, short packet transmission will be more popular and will eventually be a dominating transmission mode in machine-centric wireless systems.
We believe that the proposed SVT would serve as a useful tool in the machine communication era.

\bibliographystyle{ieeetr}
\bibliography{Reference}

\end{document}